\documentclass[aps,prl]{revtex4}

\begin{document}

\title{Dark matter as a cancer hazard}

\author{Olga~Chashchina}
\email{chashchina.olga@gmail.com}
\affiliation{\'{E}cole Polytechnique, Palaiseau, France}

\author{Zurab~Silagadze}
\email{Z.K.Silagadze@inp.nsk.su}\affiliation{ Budker Institute of 
Nuclear Physics and Novosibirsk State University, Novosibirsk 630
090, Russia }

\begin{abstract}
We comment on the paper ``Dark Matter collisions with the Human Body''
by K.~Freese and C.~Savage (Phys.\ Lett.\ B {\bf 717}, 25 (2012) 
[arXiv:1204.1339]) and describe a dark matter model for which the results of 
the previous paper do not quite apply. Within this mirror dark matter model, 
potentially hazardous objects, mirror micrometeorites, can exist and may lead 
to diseases triggered by multiple mutations, such as cancer, though with very
low probability. 
\end{abstract}

\maketitle

Is the dark matter collisions with the human body dangerous to human
health? This question was investigated and answered negatively in
the interesting article \cite{1} by Freese and Savage (possible biological
effects of dark matter were also previously discussed in \cite{1A}). 
This reassuring conclusion is based essentially on two premises. Namely, dark 
matter particles deposit much less energy in collisions 
($\sim 10~\mathrm{keV}$) than the cosmic-ray muons ($\sim 10-100\mathrm{MeV}$) 
and the expected rates of the dark matter collisions are rather low. As a 
result, dark matter related radiation exposure is negligible compared to other 
natural radiation sources and is harmless to the human body.

The quoted paper \cite{1} assumed Weakly Interacting Massive Particles (WIMPs) 
as a leading candidate for the dark matter. For WIMPs the above mentioned
two premises are well justified and the conclusions of \cite{1} are indeed 
very convincing. However, WIMPs are just one possible candidate, although most 
popular, for dark matter particles. Can the main conclusion of \cite{1} that
dark matter is harmless to human health be extended to other dark matter 
models too? Not necessarily. There is a dark matter model in which the
first premise that dark matter projectiles deposit negligible energy compared
to cosmic-ray muons might be not valid.   

Mirror dark matter with sufficient photon - mirror photon kinetic mixing 
provides the dark matter model we have in mind (for recent review and 
further references see \cite{2}). In this model the matter and gauge fields
content of the universe is doubled compared to the Standard model. Mirror 
partners of ordinary elementary particles can constitute a parallel mirror 
world as complex and rich in various structures as our own one. This 
fascinating possibility was first anticipated by Kobzarev, Okun and 
Pomeranchuk in \cite{3}. They demonstrated that besides gravity only very 
weak interactions were allowed between the mirror and ordinary particles.
Among these allowed interactions the most important for our purposes is
the possible photon - mirror photon kinetic mixing \cite{3A,3B}
\begin{equation}
{\cal L}_{mix} = \frac{\epsilon}{2} F^{\mu \nu} F'_{\mu \nu}.
\label{eq1}
\end{equation}
As a result of this mixing, mirror charged particles acquire a small ordinary
electric charge \cite{3B,4} and, in contrast to electromagnetically neutral 
WIMPs, such mirror dark matter particles scatter off ordinary nuclei via 
Rutherford-type interaction. Therefore, the reaction rates of \cite{1} for 
dark matter interactions with human body should be recalculated in the case of 
mirror dark matter. However, this is not our main concern here. We suspect 
that recalculated radiation exposure will be still negligible compared to 
other natural radiation sources in the case of mirror matter too.  What's the 
fuss then? The point is that the mirror matter provides a completely new type 
of radiation hazard not found in WIMP-type dark matter models. What are these 
new hazardous objects causing it?

Although the microphysics in the mirror world is the same as in our own one,
its macro-evolution with such key stages as baryogenesis, nucleosynthesis, 
recombination, can  proceed in somewhat different conditions than in ordinary 
world \cite{5,6}. Nevertheless, the resulting mirror world very much resembles 
our ordinary one, as far as the existence of various familiar astrophysical 
objects is concerned \cite{2,7,8}. Namely, asteroid or comet sized small 
mirror matter space bodies can exist and their collisions with the Earth can 
result in truly catastrophic events \cite{9,10}. Fortunately, it seems that 
the near Earth space is not teeming with such objects. Though, the number
of small mirror dust particles could potentially be quite large as they are
generated in collisions of mirror space bodies with themselves and 
ordinary bodies \cite{11}. The impact of mirror dust particles, mirror 
micrometeorites, on the Earth were explored in \cite{11}. In our opinion,
mirror micrometeorites constitute the above mentioned new type of 
health-hazardous objects absent in WIMP-type dark matter models. Let us take 
a closer look at how they lose their energy  when penetrating the ordinary 
matter. 

We begin by calculation of how much energy  is transferred to initially
motionless ordinary nucleus of mass $m_A$ and charge $Ze$ in Rutherford 
scattering of the mirror nucleus of mass $m_{A^\prime}$ and effective 
charge $\epsilon Z^\prime e$. We can do this as follows \cite{12}. The mirror
nucleus moving with the velocity $V$ creates the radial electric field of the
strength \cite{13}
\begin{equation}
{\cal E}=\frac{\epsilon Z^\prime e}{r^2}\,\frac{1-\beta^2}{(1-\beta^2\sin^2
{\theta})^{3/2}}
\label{eq2}
\end{equation}
at the location of the ordinary nucleus. Here $\beta=V/c$ and $\theta$ is the
angle between the relative mirror nucleus - ordinary nucleus radius-vector and
the direction of motion of the the mirror nucleus. In the Rutherford 
scattering, small scattering angles dominate and, therefore, at first 
approximation we can assume that the mirror nucleus moves at a straight line.
Then, obviously,
\begin{equation}
\sin{\theta}=\frac{b}{r},
\label{eq3}
\end{equation}
$b$ being the impact parameter, and the transverse momentum transferred to the 
ordinary nucleus during the collision is
\begin{equation}
p=\int\limits_{-\infty}^\infty Ze{\cal E}_y\,dt=\int\limits_{-\infty}^\infty 
Ze{\cal E}\,\sin{\theta}\,dt.
\label{eq4}
\end{equation}
But
\begin{equation}
dt=\frac{dx}{V}=\frac{d(b\cot{(\pi-\theta)})}{V}=\frac{b}{V}\,\frac{d\theta}
{\sin^2{\theta}},
\label{eq5}
\end{equation}
and (\ref{eq4}) takes the form
\begin{equation}
p=\frac{ZZ^\prime e^2\epsilon}{bV}\int\limits_0^\pi\frac{(1-\beta^2)
\sin{\theta}\,d\theta}{(1-\beta^2\sin^2{\theta})^{3/2}}=\frac{ZZ^\prime e^2
\epsilon}{bV}\int\limits_{-1}^1 d\frac{x}{\sqrt{1-\beta^2+\beta^2x^2}}=
\frac{2ZZ^\prime e^2\epsilon}{bV}.
\label{eq6}
\end{equation}
Therefore, the transferred energy equals to
\begin{equation}
T=\frac{p^2}{2M_A}=\frac{2}{M_A}\left (\frac{ZZ^\prime \alpha\epsilon}{bV}
\right)^2.
\label{eq7}
\end{equation}
Note that we are using natural units $\hbar=c=1$ and Gaussian units for the
electric charge, so that $e^2=\alpha$.

When a mirror micrometeorite, consisting of $N$ mirror nuclei, traverses
a distance $dx$ in the ordinary matter, the total energy deposition equals:
\begin{equation}
dE=Nn_A\,dx\int 2\pi b\, T\,|db|,
\label{eq8}
\end{equation}
where $n_A$ is the number density of ordinary nuclei. But from (\ref{eq7})
\begin{equation}
\frac{|db|}{b}=\frac{dT}{2T},
\label{eq9}
\end{equation}
and we get 
\begin{equation}
\frac{dE}{dx}=\pi Nn_A\int b^2T\frac{dT}{T}=\frac{2\pi N Z^2 Z^\prime \alpha^2
\epsilon^2 n_A}{M_AV^2}\ln{\frac{T_{max}}{T_{min}}}.
\label{eq10}
\end{equation}
The maximal energy transfer, $T_{max}$, corresponds to the head-on collision
and is equal to:
\begin{equation}
T_{max}=4T_0\frac{\mu^2_{AA^\prime}}{M_AM_{A^\prime}}=
2\frac{\mu^2_{AA^\prime}}{M_A}V^2,
\label{eq11}
\end{equation}
where $T_0=M_{A^\prime}V^2/2$ is the initial kinetic energy of the mirror 
nucleus and $\mu_{AA^\prime}=M_AM_{A^\prime}/(M_A+M_{A^\prime})$ is the 
reduced mass. 

The minimal energy transfer, $T_{min}$, can be estimated as follows.
During the collision, transverse momentum of the ordinary nucleus is uncertain
with $\Delta p_y\sim p/2$. According to the uncertainty principle,
the corresponding uncertainty in the nucleus transverse position is
$\Delta y\sim 1/\Delta p_y=2/p$. It is reasonable to require $\Delta y<r_0$,
$r_0$ being the radius at which the screening effects due to the atomic
electrons becomes effective. Therefore $p\ge 2/r_0$ and
\begin{equation}
T_{min}=\frac{1}{2M_A}\left (\frac{2}{r_0}\right )^2=\frac{2}{M_Ar_0^2}.
\label{eq12}
\end{equation}
Finally,
\begin{equation}
\frac{T_{max}}{T_{min}}=(\mu_{AA^\prime}Vr_0)^2
\label{eq13}
\end{equation}
and (\ref{eq10}) takes the form
\begin{equation}
\frac{dE}{dx}=\frac{4\pi N Z^2 Z^{\prime\,2}\alpha^2\epsilon^2 n_A}{M_AV^2}
\ln{(\mu_{AA^\prime}Vr_0)}.
\label{eq14}
\end{equation}
As far as $dE/dx$ is concerned, human body can approximately be substituted 
by water which corresponds to the following change in (\ref{eq14}):
\begin{equation}
\frac{Z^2}{M_A}n_A\,\ln{(\mu_{AA^\prime}Vr_0)}\to\frac{\rho_{H_2O}}{M_{H_2O}}
\left(2\cdot\frac{1}{u}\cdot 1.9+\frac{8^2}{16u}\cdot 4.3\right)\approx 21\,
\frac{\rho_{H_2O}}{M_{H_2O}}\,\frac{1}{u},
\label{eq15}
\end{equation}
where $u\approx 931~\mathrm{MeV}$ is the atomic mass unit and where for 
$V=30~\mathrm{km}/\mathrm{s}$, $\ln{(\mu_{HA^\prime}Vr_{0H})}\approx 1.9$ and 
$\ln{(\mu_{OA^\prime}Vr_{0O})}\approx 4.3$, if we use the 
Lindhard-Thomas-Fermi formula (which takes into account the screening effects 
from both ordinary and mirror electrons)
\cite{14}:
\begin{equation}
r_0=\frac{0.8853\,r_B}{\sqrt{Z^{2/3}+Z^{\prime\,2/3}}},\;\;
r_B\approx 5.29\cdot 10^{-9}~\mathrm{cm}\approx 2.68\cdot 10^{-4}~\mathrm{eV}
^{-1},
\label{eq16}
\end{equation}
for the effective screening radius $r_0$, and assume the mirror iron 
micrometeorite ($Z^\prime=26$). Then we get from (\ref{eq14})
\begin{equation}
\frac{dE}{dx}=\left (\frac{N}{10^{15}}\right)\left (\frac{\epsilon}
{10^{-9}}\right)^2\left (\frac{30~\mathrm{km}/\mathrm{s}}{V}\right)^2
13~\mathrm{GeV}/\mathrm{cm}.
\label{eq17}
\end{equation}
Note that (\ref{eq14}) differs by a factor of $M_{A^\prime}/M_A$
from the corresponding expression in \cite{11}. Let us explain the source
of this difference.

In one collision the mirror nucleus changes its longitudinal momentum by the 
amount 
\begin{equation}
\Delta p_x=-M_{A^\prime}V_c(1-\cos{\theta_c})\approx -M_{A^\prime}V_c
\frac{\theta_c^2}{2},
\label{eq18}
\end{equation}
where 
\begin{equation}
V_c=V-u=\frac{M_AV}{M_{A^\prime}+M_A}
\label{eq19}
\end{equation}
is the mirror nucleus velocity in the center-of-mass frame 
($u=M_{A^\prime}V/(M_{A^\prime}+M_A)$ is the velocity of the center-of-mass), 
and
\begin{equation}
\theta_c=\frac{2ZZ^\prime e^2\epsilon}{\mu_{AA^\prime}bV^2}=
\frac{2ZZ^\prime \alpha\epsilon}{M_{A^\prime}bV^2}\,\frac{M_{A^\prime}+M_A}
{M_A}
\label{eq20}
\end{equation}
is the center-of-mass scattering angle. Therefore,
\begin{equation}
\Delta p_x=-\frac{2Z^2Z^{\prime\,2}\alpha^2\epsilon^2}{b^2V^3}\,
\frac{M_{A^\prime}+M_A}{M_AM_{A^\prime}}.
\label{eq21}
\end{equation}
Note that the energy loss of the mirror nucleus after one collision equals: 
\begin{equation}
T=\frac{1}{2}M_{A^\prime}[V^2-(u+V_c\cos{\theta_c})^2-V_c^2\sin^2{\theta_c}]=
2M_{A^\prime}uV_c\sin^2{\frac{\theta_c}{2}}\approx M_{A^\prime}uV_c\frac
{\theta_c^2}{2},
\label{eq22}
\end{equation}
which is equivalent to (\ref{eq7}).

The number of collisions during a time interval $dt$ is equal to
\begin{equation}
dN_{coll}=2\pi b |db| n_A\,Vdt=\pi b^2n_AVdt\,\frac{dT}{T},
\label{eq23}
\end{equation}
and, therefore, the mirror micrometeorite loses momentum at the rate
\begin{eqnarray} &&
\frac{dP_x}{dt}=\int N\Delta p_x\,\frac{dN_{coll}}{dt}=-\frac{M_{A^\prime}+
M_A}{M_AM_{A^\prime}}\,\frac{2\pi NZ^2Z^{\prime\, 2}\alpha^2\epsilon^2n_A}
{V^2}\ln{\frac{T_{max}}{T_{min}}}= \nonumber \\ &&
-\frac{M_{A^\prime}+M_A}{M_AM_{A^\prime}}\,
\frac{4\pi NZ^2Z^{\prime\, 2}\alpha^2\epsilon^2n_A}
{V^2}\ln{(\mu_{AA^\prime}Vr_0)}.
\label{eq24}
\end{eqnarray}
For the mirror micrometeorite's kinetic energy $E_K=P_x^2/(2NM_{A^\prime})$, 
we get
\begin{equation}
\frac{dE_K}{dx}=\frac{P_x}{NM_{A^\prime}}\,\frac{dP_x}{dx}=V\frac{dP_x}{d(Vt)}
=\frac{dP_x}{dt},
\label{eq25}
\end{equation}
Therefore,
\begin{equation}
\frac{dE_K}{dx}=-\frac{M_{A^\prime}+M_A}{M_A}\,
\frac{4\pi NZ^2Z^{\prime\, 2}\alpha^2\epsilon^2\rho}
{M_AM_{A^\prime}V^2}\ln{(\mu_{AA^\prime}Vr_0)},
\label{eq26}
\end{equation}
where we have substituted $n_A=\rho/M_A$, $\rho$ being the density of the
target medium. This quantity was calculated in \cite{11} and the results
coincide if the replacement $M_A\to\mu_{AA^\prime}$ is made in the result
of \cite{11}.

Transverse momentum acquired by mirror nuclei in the Rutherford scattering is
dissipated as heat $Q$ in the mirror micrometeorite whose temperature will
rise. This circumstance explains the difference between (\ref{eq26}) and
(\ref{eq14}). Namely,
\begin{equation}
\frac{dE}{dx}=\frac{M_{A^\prime}}{M_A+M_{A^\prime}}\left (-\frac{dE_K}{dx}
\right ),\;\;\; \frac{dQ}{dx}=\frac{M_A}{M_A+M_{A^\prime}}\left (-\frac{dE_K}
{dx}\right ).
\label{eq27}
\end{equation}
For water and mirror iron micrometeorite, $M_{A^\prime}$ is significantly 
larger than $M_A$ and in the first approximation the difference between
$dE_K/dx$ and $-dE/dx$ can be neglected. Then we can take $E=NM_{A^\prime}V^2
/2$ in (\ref{eq17}) and solve the resulting differential equation for $V(x)$.
As a result we get that the stopping distance $L$ in water for the iron mirror 
micrometeorite moving with the initial velocity $V$ can be estimated 
as:
\begin{equation}
L=\left(\frac{10^{-9}}{\epsilon}\right)^2\left(\frac{V}{30~\mathrm{km}/
\mathrm{s}}\right)^4 100~\mathrm{km}.
\label{eq28}
\end{equation}
It was argued \cite{2} that the mirror dark matter can provide an adequate 
description of the known dark matter phenomena provided that $\epsilon\sim
10^{-9}$. Then two important conclusions can be drawn from (\ref{eq28})
and (\ref{eq17}). The first indicates that mirror micrometeorites reach the 
Earth's surface essentially without losing their velocity in the atmosphere 
for all range of expected initial velocities $11-70~\mathrm{km}/\mathrm{s}$ 
(for mirror micrometeorites which are bound to the solar system and which
have not yet been trapped by the Earth). On the other hand, according 
to (\ref{eq17}), when moving through a human body they deposit a lot of energy 
greatly exceeding energy deposition from cosmic-ray muons. It is true, however,
that in the case of the mirror micrometeorite, the energy deposition doesn't 
have a point-like character thus involving many and many target molecules. 
The second difference from the cosmic-ray muons is that energy deposition from 
the  mirror micrometeorite is not ionizing. The cosmic-ray muons move with 
velocities much greater than the velocities of atomic electrons. In this case 
Rutherford scattering on atomic electrons dominates and leads to ionization 
energy losses well described by the celebrated Bethe-Bloch formula \cite{15}. 
Energy losses due to Rutherford scattering on target nuclei leading to the 
corresponding nuclear recoils are negligible. On the contrary, mirror 
micrometeorite's velocities are much smaller than the atomic electrons 
velocities, and in this case energy losses due to the Rutherford scattering on 
target nuclei dominates while the scattering effects on electrons give 
a negligible contribution \cite{15}.

Transferred energy (\ref{eq17}) will be dissipated as vibrations and 
rearrangements of the target biological molecules. Each mirror nuclei
will collide with $\sigma n_A$ ordinary nuclei per length unit. Therefore,
the total number of collisions per unit length is
\begin{equation}
\frac{dN_{coll}}{dx}=N\sigma n_A.
\label{eq29}
\end{equation}
Here $\sigma$ is the screened Rutherford cross section which can be obtained
by integrating the differential cross section \cite{15A}
\begin{equation}
\frac{d\sigma}{d\Omega}=\frac{Z^2Z^{\prime\,2}\epsilon^2\alpha^2}
{\mu^2_{AA^\prime}V^4(1-\cos{\theta}+\alpha_0)^2},\;\;\;
\alpha_0=\frac{1}{2(\mu_{AA^\prime}Vr_0)^2}.
\label{eq30}
\end{equation}
The result is \cite{15A}
\begin{equation}
\sigma=\frac{4\pi Z^2Z^{\prime\,2}\epsilon^2\alpha^2}{\mu^2_{AA^\prime}V^4
\alpha_0(2+\alpha_0)}\approx \frac{4\pi Z^2Z^{\prime\,2}\epsilon^2\alpha^2}
{V^2}\,r_0^2,
\label{eq31}
\end{equation}
because $\alpha_0\ll 1$.

For sufficiently large momentum transfer, nucleus structure becomes relevant
and we should include nucleus form factor in the cross section calculation 
to take into account the corresponding decoherence effects. In our case,
however, simple estimates show that the decoherence effects are irrelevant.
Indeed, only for the Milky Way bound dark matter with $V\sim 300~\mathrm{km}/
\mathrm{s}$, when the typical momentum transfer is \cite{15B} $q=M_{A^\prime}
|\Delta \vec{V}|\sim 10^{-3}M_{A^\prime}\approx 50~\mathrm{MeV}\approx
(4~\mathrm{fm})^{-1}$, $q^{-1}\sim 4~\mathrm{fm}$ gets comparable to the size 
of mirror iron nucleus $R_{A^\prime}=1.2A^{\prime\,1/3}~\mathrm{fm}\approx 
4.6~\mathrm{fm}$, and we should consider (presumably still rather small) 
decoherence effects. For $V\sim 70~\mathrm{km}/\mathrm{s}$ (maximal Earth 
impact velocity for solar bound dark matter at Earth's location), the typical 
momentum transfer will be four times smaller and thus, decoherence effects can 
be safely ignored.

Substituting (\ref{eq31}) into (\ref{eq29}), we get
\begin{equation}
\frac{dN_{coll}}{dx}=\frac{4\pi NZ^2Z^{\prime\,2}\epsilon^2\alpha^2n_A}
{V^2}\,r_0^2.
\label{eq32}
\end{equation}
Therefore, the average energy transfer per interaction equals to
\begin{equation}
\epsilon=\left .\left(\frac{dE}{dx}\right)\right /\left(\frac{dN_{coll}}{dx}
\right)=\frac{\ln{(\mu_{AA^\prime}Vr_0)}}{M_Ar_0^2}.
\label{eq33}
\end{equation}
This quantity depends very weakly on the velocity $V$ and for hydrogen,
oxygen, nitrogen and carbon targets, assuming $V=30~\mathrm{km}/\mathrm{s}$ 
and mirror iron micrometeorite, we get $\epsilon_H\approx 0.4~\mathrm{eV}$,
$\epsilon_O\approx 0.07~\mathrm{eV}$,  $\epsilon_N\approx 0.07~\mathrm{eV}$
and $\epsilon_C\approx 0.08~\mathrm{eV}$ respectively. Hence, the average value
for the water target is $\epsilon_{H_2O}=(2\epsilon_H+\epsilon_O)/3\approx
0.3~\mathrm{eV}$.

Let us estimate how much the atoms of human DNA could be affected by 
interactions with mirror iron micrometeorite. For a given atom in the DNA the 
number of collisions with mirror nuclei equals to $p\sim\sigma R n_{A^\prime}$,
where
\begin{equation}
R=\left(\frac{3N}{4\pi n_{A^\prime}}\right )^{1/3}\approx 1.3\cdot 10^{-3}
\left(\frac{N}{10^{15}}\right)^{1/3}~\mathrm{cm}
\label{eq34}
\end{equation} 
is the size of the micrometeorite assuming it has spherical form and
taking $n_{A^\prime}\sim 10^{23}~\mathrm{cm}^{-3}$. Using equations
(\ref{eq17}), (\ref{eq29}), (\ref{eq33}) and assuming  $n_{A^\prime}\sim
n_A$, $\epsilon\sim 0.3~\mathrm{eV}$, we can estimate $p$ as follows:
\begin{equation}
p\sim\frac{n_{A^\prime}}{n_A}\frac{1}{N}\frac{dN_{coll}}{dx}\,R\sim
\frac{1}{N\epsilon}\frac{dE}{dx}\,R\approx 6\cdot 10^{-8}
\left (\frac{N}{10^{15}}\right)^{1/3}\left (\frac{\epsilon}
{10^{-9}}\right)^2\left (\frac{30~\mathrm{km}/\mathrm{s}}{V}\right)^2.
\label{eq35}
\end{equation}
As we see, $p$ is a very small number. In fact, it gives the probability that
a given ordinary atom will be involved in the interaction when a mirror
micrometeorite passes through it. However, there are about $N_{DNA}\sim 
10^{11}$ atoms in the human DNA (to quote Carl Sagan's apt comparison,
``There are as many atoms in one molecule of DNA as there are stars in a 
typical galaxy''). Therefore, when a mirror micrometeorite passes through a
human DNA, the expected number of perturbed atoms can be estimated as
\begin{equation}
pN_{DNA}\sim 6\cdot 10^3
\left (\frac{N}{10^{15}}\right)^{1/3}\left (\frac{\epsilon}
{10^{-9}}\right)^2\left (\frac{30~\mathrm{km}/\mathrm{s}}{V}\right)^2.
\label{eq36}
\end{equation}
However, not all collisions can lead to mutations but only those in which
energy transfer exceeds some threshold value $E_{mut}$. From eq.(\ref{eq22})
we get that in this case the center-of-mass scattering angle $\theta_c$
satisfies
\begin{equation}
\theta_c>\theta_{c0}=\frac{\sqrt{2M_A\,E_{mut}}}{\mu_{AA^\prime}V}.
\label{eq37}
\end{equation}
Correspondingly, the angular integration range for the equation (\ref{eq30})
is reduced, and we get
\begin{equation}
\sigma_{mut}=\frac{\alpha_0}{2}\,\frac{2+\theta^2_{c0}/2}{\alpha_0+
\theta^2_{c0}/2}\,\sigma\approx \frac{2\alpha_0}{\theta^2_{c0}}\,\sigma,
\label{eq38}
\end{equation}
as $\alpha_0\ll\theta^2_{c0}/2\ll 1$ for the relevant range of parameters.
The suppression factor
\begin{equation}
\eta=\frac{2\alpha_0}{\theta^2_{c0}}=\frac{1}{2M_A E_{mut} r_0^2},
\label{eq39}
\end{equation}
assuming $E_{mut}=1~\mathrm{eV}$ and mirror iron micrometeorite, equals to 
$\eta_H\approx 9\cdot 10^{-2}$, $\eta_O\approx 7.6\cdot 10^{-3}$, 
$\eta_C\approx 9.6\cdot 10^{-3}$ and $\eta_N\approx 8.5\cdot 10^{-3}$ 
respectively for  hydrogen, oxygen, carbon and nitrogen. As atoms of these 
elements are present in DNA roughly in comparable amounts, for estimate we use 
the avaraged suppression factor $\eta=(\eta_H+\eta_O+\eta_C+\eta_N)/4
\approx 3\cdot 10^{-2}$. As a result, we get the following order of magnitude 
estimate of the number of mutations in DNA:
\begin{equation}
N_{mut}\sim 180 \left (\frac{1~\mathrm{eV}}{E_{mut}}\right)
\left (\frac{N}{10^{15}}\right)^{1/3}\left (\frac{\epsilon}
{10^{-9}}\right)^2\left (\frac{30~\mathrm{km}/\mathrm{s}}{V}\right)^2.
\label{eq40}
\end{equation}
We can thus speculate that the mirror micrometeorite, when interacting with 
the DNA molecules, can lead to multiple simultaneous mutations and potentially 
cause disease (note that the energy of only 0.1--10~eV is required to displace 
an atom in organic molecules and cause a DNA strand break \cite{1A}). 
There is an evidence that individual malignant cancer cells in human tumors 
contain thousands of random mutations and that to account for these multiple
mutations rates found in human cancers it is necessary to assume that the 
usual mechanisms to repair corrupted DNA somehow were also damaged.
It was suggested, therefore, that mutation accumulation during tumor 
progression due to lesion of DNA repair mechanisms probably plays a major role 
in triggering the cancer growth \cite{16}. Thus, it can turn out that mirror 
micrometeorites are much more dangerous carcinogens than other natural 
radiation sources because of their potential capability to produce 
simultaneous multiple mutations and at the same time damage genes that control 
DNA repair mechanisms. It is even not excluded that the passage of solar 
system through a dense mirror dust cloud can lead to mass extinctions 
\cite{17A}. However, we suspect that usually the probability that the DNA 
damage due to mirror micrometeorite eventually will lead to cancer is very low 
because even hundred mutations, as a rule, is insufficient to significantly 
deteriorate multiple pathways for DNA repair found in normal cells. 

On the other hand, multiple simultaneous mutations may allow organisms to leap 
across fitness valleys and thus can potentially be beneficial for evolution  
\cite{17}. It is possible, therefore, that mirror micrometeorites, if 
proven to exist, had played a role in rare evolutionary events requiring 
simultaneous multiple mutations (potential role of  mutagenic effects 
of dark matter in observed diversification of life after mass extinctions
was also suggested in \cite{17A}).

To conclude, we have indicated a dark matter model for which the conclusion
of \cite{1} that the dark matter is harmless for human health, doesn't 
directly apply. It will be premature to worry about though. Although some 
cosmological, astrophysical and experimental facts can be interpreted in favor 
of mirror dark matter existence \cite{2,18,19}, no definite conclusions can be 
drawn at present and mirror dark matter, not to speak of mirror 
micrometeorites, remains a highly speculative idea. Besides, even if we assume 
the reality of this threat, we have every reason to consider the health risks 
as normally very low (excluding, perhaps, such rare speculative events as 
passing of the solar system through a dense mirror dust cloud mentioned above) 
because all living organisms have been continuously exposed to this threat 
from the dawn of life and had plenty of time to mitigate this risk factor.

Moreover, normally the flux of mirror micrometeorites is not expected to be
large. It is known \cite{20,21} that the flux of ordinary micrometeorites
before atmospheric entry is about $3\cdot 10^4$ tons per year. It is difficult
to reliably estimate the amount of mirror matter in the solar system but a 
very rough estimate indicates that about $10^{-5}$ fraction of all solar 
system matter could be mirror matter which might have accumulated in the 
vicinity of the solar system during its formation \cite{2}. Therefore, we will 
assume that the flux of mirror micrometeorites doesn't exceed 300 kg per year. 
Assuming that mirror micrometeorites contain $10^{15}$ mirror iron nuclei, 
this number corresponds to about $3\cdot 10^{12}$ mirror micrometeorites per 
year. Thus, the probability that one of these micrometeorites hits a human 
with about m$^2$ effective cross section  doesn't exceed 
$3\cdot 10^{12}/(4\pi R^2_{Earth})\approx 6\cdot 10^{-3}$ per year, which is 
quite a small number.
 
Mirror dust particles can potentially be observed in cryogenic detectors such 
as NAUTILUS gravitational wave detector \cite{11}. Interestingly, NAUTILUS has 
found several anomalously large energy depositing events which if interpreted 
as caused by mirror dust particles imply the flux of about 
$2\cdot10^{-6}~\mbox{m}^{-2}\mbox{s}^{-1}$ \cite{11}. Such a flux corresponds 
to several tens of mirror micrometeorites impact events per year and per human 
and is several orders of magnitude larger than the flux estimated above. 
It seems unrealistic that these NAUTILUS events are due to mirror 
micrometeorites impacts but we hope that our observation that such impacts 
might be hazardous to human health will stimulate experimental searches of 
mirror micrometeorites and further dark matter research. 

\section*{Acknowledgments}
We thank Konstantin Zioutas for informing us about his 1990 paper \cite{1A},
as well as Robert Foot and Sabine Hossenfelder for helpful comments.
We are grateful to the anonymous Reviewer for the constructive critical remarks
which helped us to improve and make more clear our initially rather imperfect
presentation. The work of Z.K.S. is supported by the Ministry of Education 
and Science of the Russian Federation and in part by Russian Federation 
President Grant for the support of scientific schools NSh-8367.2016.2.

\end{document}